\documentclass[floatfix,twocolumn,aps,prd,showpacs,amsmath,amssymb]{revtex4}
\usepackage{epsfig}

\usepackage[latin1]{inputenc}
\begin{document}

\title{Exact solution for the energy density inside a one-dimensional non-static cavity with an arbitrary initial field state}

\author{Danilo T. Alves$^{1}$, Edney R. Granhen$^{2}$, Hector O. Silva$^{1}$ and Mateus G. Lima$^{1}$}
\affiliation{(1) - Faculdade de F\'\i sica, Universidade Federal do
Par\'a, 66075-110, Bel\'em, PA,  Brazil\\(2) - Centro Brasileiro de Pesquisas Físicas, Rua Dr. Xavier Sigaud,
150, 22290-180, Rio de Janeiro, RJ, Brazil}

\date{\today}
\begin{abstract}
We study the exact solution for the energy density of
a real massless scalar field in a two-dimensional spacetime, inside a non-static cavity
with an arbitrary initial field state, taking into account the Neumann and Dirichlet boundary conditions. 
This work generalizes the exact
solution proposed by Cole and Schieve in the context
of the Dirichlet boundary condition and vacuum as the initial state. 
We investigate diagonal states, examining 
the vacuum and thermal field as particular cases.
We also study non-diagonal initial field states,
taking as examples the coherent and Schrödinger cat states.
\end{abstract}

\maketitle

\section{Introduction}

The first works investigating the
quantum problem of the radiation generated by 
moving mirrors in vacuum where published in the 1970s decade 
(see Refs. \cite{Moore-1970, Fulling-Davies-PRS-1976-I, 
Fulling-Davies-PRS-1977-II, trabalhos-pioneiros}). 
Moore \cite{Moore-1970}, in the context of a real massless scalar field in a two dimensional spacetime, 
investigated the radiation generated in a cavity with a moving boundary. Imposing
Dirichlet boundary condition to the field and also a prescribed law for the movement
of a boundary, Moore obtained an exact formula for the expected value of the energy-momentum tensor,
assuming the initial field state as the vacuum. The field solution obtained by Moore
is given in terms of a functional equation, usually called Moore's equation
(which was obtained independently by Vesnitskii \cite{Vesnitskii-1972}),
for which there is no general technique of analytical solution. 
Law \cite{Law-PRL-94} obtained an exact analytic solution for the Moore equation for a particular resonant 
movement of the boundary (see also Ref. \cite{solucoes-analiticas-exatas}). Cole and Schieve \cite{Cole-Schieve-1995} 
proposed a numerical method to solve exactly the Moore equation for a general law of motion
of the boundary. Approximate analytical solutions of the Moore equation were also obtained, for instance, by Dodonov-Klimov-Nikonov 
\cite{Dodonov-JMP-1993} and Dalvit-Mazzitelli \cite{Dalvit-PRA-1998}.
With different approaches from those adopted by Moore \cite{Moore-1970}
and Fulling-Davies \cite{Fulling-Davies-PRS-1976-I}, 
perturbative methods where developed to solve the problem
of a quantum field in the presence of a single moving boundary
\cite{perturbative-approach-one-boundary} and also in oscillating cavities 
\cite{perturbative-approach-cavities}.

The first works studying the problem of particle creation
by moving mirrors with initial states different from 
vacuum were also published about thirty years ago \cite{Fulling-Davies-PRS-1977-II},
showing that the presence of real particles in the initial 
state amplifies the phenomenon of particle creation.
Considering a thermal bath as the ``in" field state, the dynamical Casimir effect
has been investigated for the case of a single mirror \cite{temperatura-uma-fronteira,plunien-PRL-2000,alves-granhen-lima-PRD-2008},
and also for an oscillating cavity \cite{Dodonov-JMP-1993,plunien-PRL-2000,temperatura-cavidade}.
The coherent state as initial field state has been considered for one
moving boundary in Refs. \cite{alves-granhen-lima-PRD-2008, Alves-Farina-Maia-Neto-JPA-2003, estados-coerentes,estados-coerentes-2}, and the superposition of coherent sates in connection to the investigation of decoherence via the dynamical 
Casimir effect \cite{estados-coerentes-superpostos}. Squeezed states have also been considered 
by three of the present authors \cite{alves-granhen-lima-PRD-2008}.

Recently, several works have also investigated the influence of different boundary conditions on 
the dynamical Casimir effect \cite{alves-granhen-lima-PRD-2008,cas-din-papel-conds-fronteira,Alves-Granhen-2008,Alves-Farina-Maia-Neto-JPA-2003}.
In this context, it was showed that, for the single mirror problem, Dirichlet and Neumann boundary 
conditions yield the same energy density radiated when the initial field state 
is symmetrical under time translations \cite{alves-granhen-lima-PRD-2008,
Alves-Farina-Maia-Neto-JPA-2003}. 

In the present work we investigate the time evolution of the energy density for
a real massless scalar field in a two-dimensional spacetime, inside a non-static cavity,
considering both the Dirichlet and Neumann boundary conditions. 
We extend to an arbitrary initial field state the exact method for calculating the energy density
proposed by Cole and Schieve \cite{Cole-Schieve-1995,Cole-Schieve-2001} in the context of the Dirichlet boundary condition and vacuum as the initial state. Considering an arbitrary initial field state, we show
that the energy density in a given point of the spacetime can be obtained
by tracing back a sequence of null lines, connecting the value of the energy density 
at the given spacetime point to a certain known value of the energy at a 
point in the ``static zone'', where the initial field modes
are not affected by the perturbation caused by the boundary motion. 
We investigate diagonal states (for which the static energy density
is invariant under time translation), examining in particular
the vacuum and thermal field states.
We also study non-diagonal initial field states
(in this case the static energy density
is not invariant under time translation),
investigating the coherent and the Schrödinger cat states.

The paper is organized as follows. In Sec. \ref{exact-formulas},
considering Neumann and Dirichlet boundary conditions and also an arbitrary initial field state, we write 
the field solution and the energy density in the non-static cavity, generalizing the formula found in the literature \cite{Moore-1970,Fulling-Davies-PRS-1976-I}, 
which is valid for the Dirichlet boundary condition and vacuum as the initial field state. 
In Sec. \ref{static-situation} we discuss the energy density
in the static situation. In Sec. \ref{non-static-situation} we
obtain the energy density in the non-static situation given in terms of the static energy density.
In Sec. \ref{final-comments} we summarize our main results.


\section{Energy density: exact formulas}
\label{exact-formulas}
Let us start considering the field satisfying the Klein-Gordon equation
(we assume throughout this paper $\hbar=c=k_B=1$):
$
\left(\partial _{t}^{2}-\partial _{x}^{2}\right) \psi \left(
t,x\right) =0,
$
and obeying conditions imposed at the static boundary located at $x=0$,
and also at the moving boundary's position at $x=L(t)$, where $L(t)$ is a prescribed law
for the moving boundary with $L(t<0)=L_0$,
where $L_0$ is the length of the cavity in the static situation.
We consider four types of boundary conditions. 
The Dirichlet-Neumann (DN) boundary condition imposes Dirichlet condition at the static boundary,
whereas the space derivative of the field
taken in the instantaneously co-moving Lorentz frame 
vanishes at the moving boundary's position (Neumann condition):
$
\partial _{x^{\prime }}\psi ^{\prime }(t^{\prime},x^{\prime})\vert_{boundary}=0\text{.}
$
Using the appropriate Lorentz transformation, this boundary condition 
can be written in terms of quantities in the laboratory inertial frame as follows:
{\small
$
\partial _{x^{\prime }}\psi ^{\prime }\left( t^{\prime },x^{\prime }\right)
= \left. \left[ \left( \dot{L}(t){\partial_t}+{\partial_x}\right) \psi \left( t,x\right) \right] \right|_{x=L(t)}=0.
$}
The Neumann-Neumann (NN) boundary condition imposes 
{\small
$
\left.\left[\partial _{x}\psi\left( t,x\right)\right]\right|_{x=0}=0
$}
and
{\small
$
\left. \left[ \left( \dot{L}(t){\partial_t}+{\partial_x}\right) \psi \left( t,x\right) \right] \right|_{x=L(t)}=0.
$}
The Neumann-Dirichlet (ND) boundary condition imposes 
{\small
$
\left.\left[\partial _{x}\psi\left( t,x\right)\right]\right|_{x=0}=0
$}
and
{\small
$
\psi\left( t,L(t)\right)=0,
$}
whereas the Dirichlet-Dirichlet (DD) boundary condition imposes
{\small
$
\psi\left( t,0\right)=0,
$}
and
{\small
$
\psi\left( t,L(t)\right)=0.
$}
Considering the procedure adopted in Refs. \cite{Moore-1970, Fulling-Davies-PRS-1976-I}, 
the field in the cavity can be obtained
by exploiting the conformal invariance of the Klein-Gordon equation. 
The field operator, solution of the wave equation, is given by:
\begin{equation}
\hat{\psi}(t,x)=\lambda[\hat{A}+\hat{B}\psi^{(0)}\left( t,x\right)]+\sum^{\infty }_{n=1-2\beta}\left[
\hat{{a}}_{n}\psi_{n}\left( t,x\right) +H.c.\right],
\label{field-solution-1}
\end{equation} 
where $\psi^{(0)}=\left[R(v)+R(u)\right]/2$ \cite{Dalvit-JPA-2006}, 
and the field modes $\psi_{n}(t,x)$ are given by:
\begin{eqnarray}
\psi_{n}(t,x)=\frac{1}{\sqrt{4(n+\beta)\pi}}\left[\gamma\;\varphi_{n}^{(\beta)}(v)
+ \gamma ^{*}\;\varphi_{n}^{(\beta)}(u)\right],
\label{field-solution-2}
\end{eqnarray} 
with $\varphi_{n}^{(\beta)}(z)=e^{-i(n+\beta)\pi R(z)}$, $u=t-x$, $v=t+x$, and $R$ satisfying Moore's functional equation:
\begin{equation}
R[t+L(t)]-R[t-L(t)]=2.
\label{Moore-equation}
\end{equation}
The operators
$\hat{A}$ and $\hat{B}$ satisfy the commutation rules 
$\left[\hat{A},\hat{B}\right]=i$, 
$\left[\hat{A},\hat{a}_{n}\right]=\left[\hat{B},\hat{a}_{n}\right]=0$ \cite{Dalvit-JPA-2006}.    
In Eqs. (\ref{field-solution-1}) and (\ref{field-solution-2}) we introduce a notation
which enables us to put into a single formula the solutions for the four
boundary conditions considered in the present work. In this sense, 
for $\lambda=\gamma=1$ and $\beta=0$, Eqs. (\ref{field-solution-1}) and (\ref{field-solution-2}) 
give the NN solution. 
The other three possible situations are recovered if we consider $\lambda=0$ and: $\beta=0$ and $\gamma=i$ for the DD case;
$\beta=1/2$ and $\gamma=i$ for the DN case; $\beta=1/2$ and  $\gamma=1$ for the ND case. 

Heareafter, considering the Heisenberg picture, we are interested in 
the averages $\langle...\rangle$ taken over any initial field state 
annihilated by $\hat{B}$. In this context, we will write the exact formulas 
for the expected value of the energy 
density operator ${\cal T}=\langle\hat{T}_{00}(t,x)\rangle$. 
We can split ${\cal T}$, writing:
\begin{equation}
{\cal T}={\cal T}_{\mbox{\footnotesize vac}}+ {\cal T}_{\mbox{\footnotesize non-vac}}, 
\label{T}
\end{equation}
where
\begin{equation}
{\cal T}_{\;\mbox{\footnotesize vac}}=\frac{\pi{\left|\gamma\right|}^{2}}{4}\sum^{\infty}_{n=1-2\beta}(n+\beta)\left[{R^\prime}^{2}(v)+{R^\prime}^{2}(u)\right]
\label{vac}
\end{equation}
and 
\begin{equation}
{\cal T}_{\;\mbox{\footnotesize non-vac}}={\cal T}_{\left\langle \hat{a}^{\dag}\hat{a}\right\rangle}+
{\cal T}_{\left\langle \hat{a}\hat{a}\right\rangle},
\end{equation} 
with
\begin{equation}
{\cal T}_{\left\langle \hat{a}^{\dag}\hat{a}\right\rangle}
=g_{1}(v) + g_{1}(u),
\end{equation}
\begin{equation}
{\cal T}_{\left\langle \hat{a}\hat{a}\right\rangle}=g_{2}(v) + g_{2}(u),
\end{equation}
\begin{eqnarray}
g_{1}(z) &=& \frac{\pi\left\vert \gamma \right\vert ^{2} }{2}\sum_{n,n^{\prime}=1-2\beta}^{\infty }
\sqrt{\left( n+\beta \right) \left(
n^{\prime }+\beta \right) }  
\nonumber  \\
&&
\times\; {\mbox Re} \left\{e^{i\left(
n-n^{\prime }\right) \pi R\left(z\right) }\left[R^{\prime }\left(z\right)
\right]^{2} \left\langle \hat{a}_{n}^{\dag }\hat{a}_{n^{\prime }}\right\rangle\right\},
\label{g1}
\end{eqnarray}
\begin{eqnarray}
g_{2}(z) &=&-\frac{\pi\gamma ^{2} }{2}\sum_{n,n^{\prime }=1-2\beta}^{\infty }\sqrt{\left( n+\beta \right) \left( n^{\prime }+\beta
\right) }  \nonumber \\
&&\times\;{\mbox Re} \left\{e^{-i\left( n+n^{\prime }+2\beta
\right) \pi R\left( z\right) }\left[ R^{\prime }\left( z\right) \right]
^{2}\right.   \nonumber \\
&&\left. \times \left\langle \hat{a}_{n}\hat{a}_{n^{\prime }}\right\rangle
\right\}.   
\label{g2}
\end{eqnarray}
The term ${\cal T}_{\;\mbox{\footnotesize vac}}$  is the local energy density related to the vacuum state, which is divergent. 
Following Ref. \cite{Fulling-Davies-PRS-1976-I}, we adopt the point-splitting regularization method and obtain
${\cal T}_{\;\mbox{\footnotesize vac}}$ (now redefined) as the renormalized local energy density:
\begin{equation}
{\cal T}_{\;\mbox{\footnotesize vac}} = -f(v) -f(u), 
\label{T-vac-ren}
\end{equation}
where
\begin{equation}
f=\frac{|\gamma|^2}{24\pi}\left\{\frac{R^{\prime\prime\prime}}{R^{\prime}}-\frac{3}
{2}\left(\frac{R^{\prime\prime}}{R^{\prime}}\right)^{2}+\pi
^{2}\left[\frac{1}{2}-3(\beta-\beta^2)\right]{R^{\prime}}^{2}\right\}. 
\label{T-vac-f}
\end{equation}
In the above equation the derivatives are taken with respect to the argument of the $R$ function.

We can also write the non-vacuum part of the energy density 
in the analogous notation:
\begin{equation}
{\cal T}_{\;\mbox{\footnotesize non-vac}}=-g(v)-g(u), 
\label{T-nov-vac-g}
\end{equation}
where $g=-(g_1+g_2)$. Note 
that ${\cal T}_{\;\mbox{\footnotesize vac}}$ and ${\cal T}_{\left\langle \hat{a}^{\dag}\hat{a}\right\rangle}$ 
depend on $|\gamma|^2$, which has the same value for the considered boundary
conditions. On the other hand $
{\cal T}_{\left\langle \hat{a}\hat{a}\right\rangle}$
depends on $\gamma^{2}$, which can differ by a sign, depending
on the situation. For further analysis, it is useful to write:
\begin{equation}
{\cal T}=-h(v)-h(u), 
\label{T-em-termos-h}
\end{equation}
where
\begin{equation}
h=f+g.
\end{equation}
%
\section{Static situation}
\label{static-situation}
Let us first examine the formulas in the preceding section for the static situation $t\leq 0$,
when the Moore equation is reduced to $R(t+L_0)-R(t-L_0)=2\;$.
For this case the function $R$ is given by \cite{Moore-1970}:
\begin{equation}
R(z)={z}/{L_0}.
\label{R-static}
\end{equation}
The functions
$f$, $g$, $g_1$ and $g_2$, now relabeled, respectively, as
$f^{(s)}$, $g^{(s)}$, $g_1^{(s)}$ and $g_2^{(s)}$, are given by:
\begin{equation}
f^{(s)}=\frac{{|\gamma|^2}\pi}{{24}L_0^2}\left[\frac{1}{2}-3(\beta-\beta^2)\right],
\label{f-static}
\end{equation}
\begin{equation}
g^{(s)}=-(g_1^{(s)}+g_2^{(s)}),
\label{g-static}
\end{equation}
\begin{eqnarray}
g_1^{(s)}(z) &=& \frac{\pi\left\vert \gamma \right\vert ^{2} }{2L_0^2}\sum_{n,n^{\prime}=1-2\beta}^{\infty }\sqrt{\left( n+\beta \right) \left(
n^{\prime }+\beta \right) }  
\nonumber  \\
&&
\times\; {\mbox Re}\left\{e^{i\left(
n-n^{\prime }\right) \pi z/L_0 }
\left\langle \hat{a}_{n}^{\dag }\hat{a}_{n^{\prime }}\right\rangle\right\},   
\label{g1-static}
\end{eqnarray}
\begin{eqnarray}
g_2^{(s)}(z) &=&-\frac{\pi\gamma ^{2} }{2L_{0}^{2}}
\sum_{n,n^{\prime }=1-2\beta}^{\infty }%
\sqrt{\left( n+\beta \right) \left( n^{\prime }+\beta \right) }
\nonumber \\
&&\times\;{\mbox Re}\left\{e^{-i\left( n+n^{\prime }+2\beta
\right) \pi z/L_{0} }
\left\langle \hat{a}_{n}\hat{a}_{n^{\prime }}\right\rangle
\right\}.   
\label{g2-static}
\end{eqnarray}
Note that $f^{(s)}$ is a constant 
and ${\cal T}_{\;\mbox{\footnotesize vac}}$ (in this
case the Casimir energy density and relabeled as ${\cal T}_{\;\mbox{\footnotesize cas}}$)
is such that (see Refs. \cite{Dalvit-JPA-2006,Boyer-AJP-2003}):
\begin{eqnarray}
{\cal T}_{\;\mbox{\footnotesize cas}}^{\;\mbox{\footnotesize DD}}={\cal T}_{\;\mbox{\footnotesize cas}}^{\;\mbox{\footnotesize NN}}
=-\frac{\pi}{24L_0^2},
\;{\cal T}_{\;\mbox{\footnotesize cas}}^{\;\mbox{\footnotesize DN}}={\cal T}_{\;\mbox{\footnotesize cas}}^{\;\mbox{\footnotesize ND}}
=\frac{\pi}{48L_0^2},
\label{T-cas}
\end{eqnarray}
where the superscripts DD, NN, DN and ND mean the types of boundary conditions considered in the calculations. Note that, for mixed boundary
conditions (in this case ND or DN),
the Casimir energy is positive, originating a repulsive Casimir force (see Ref. \cite{Boyer-PRA-1974}).
On the other hand, $g^{(s)}$ is, in general, spacetime dependent,
what implies that ${\cal T}_{\left\langle \hat{a}^{\dag}\hat{a}\right\rangle}$ and 
${\cal T}_{\left\langle \hat{a}\hat{a}\right\rangle}$, in this static situation
relabeled respectively as  ${\cal T}_{\left\langle \hat{a}^{\dag}\hat{a}\right\rangle}^{(s)}$ and 
${\cal T}_{\left\langle \hat{a}\hat{a}\right\rangle}^{(s)}$, are functions of the spacetime variables. 
Despite the fact that ${\cal T}_{\left\langle \hat{a}^{\dag}\hat{a}\right\rangle}^{(s)}$ and 
${\cal T}_{\left\langle \hat{a}\hat{a}\right\rangle}^{(s)}$ can depend on time, from the principle
of energy conservation the total energy is a constant in time for the static situation.
In fact, for an arbitrary initial field state, the integration of Eqs. (\ref{g1-static}) and (\ref{g2-static}) yields:
\begin{eqnarray}
\int_0^{L_0}{\cal T}_{\left\langle \hat{a}^{\dag}\hat{a}\right\rangle}^{(s)}\;dx&=&\sum_{n=1-2\beta}^{\infty }\omega_{n}{\cal N}_n,   
\label{int-g1}
\\
\int_0^{L_0}{\cal T}_{\left\langle \hat{a}\hat{a}\right\rangle}^{(s)}\;dx&=&0,  
\label{int-g2}
\end{eqnarray}
where $\omega_{n}=\pi(n+\beta)/L_0$, and
${\cal N}_n=\left\langle \hat{a}_{n}^{\dag }\hat{a}_{n}\right\rangle$ is the initial 
mean number of particles in the $n$th mode.
Then, for the static cavity, the function ${\cal T}_{\left\langle \hat{a}\hat{a}\right\rangle}^{(s)}$
can give contribution for the spacetime behavior of the energy density $\cal T$ (now relabeled as
${\cal{T}}^{(s)}$), but not for the total energy stored in the cavity. The total
static energy ${\cal E}^{(s)}=
\int_0^{L_0}{\cal{T}}^{(s)}dx$ can be written as ${\cal E}^{(s)}={\cal E}_{\mbox{\footnotesize cas}}+ \sum_{n}\omega_{n}{\cal N}_n$,
where ${\cal E}_{\mbox{\footnotesize cas}}=\int_0^{L_0} {\cal{T}}_{\;\mbox{\footnotesize cas}}\;dx$.
Observe that, for a static cavity, for any initial field state considered, we have: ${\cal E}^{(s)\mbox{\footnotesize ND}}={\cal E}^{(s)\mbox{\footnotesize DN}}$ and ${\cal E}^{(s)\mbox{\footnotesize DD}}={\cal E}^{(s)\mbox{\footnotesize NN}}$.

For the initial field states such that the density matrix is diagonal in the Fock basis,
so that $\left\langle \hat{a}_{n}\hat{a}_{n^{\prime }}\right\rangle
=0$ and $\left\langle \hat{a}_{n}^{\dag }\hat{a}_{n^{\prime }}\right\rangle={\cal N}_{n}{\delta}_{nn^{\prime}}$
\cite{estados-coerentes-2},
we have:
\begin{equation}
g^{(s)}_{2}(z)= 0,\;\partial_z[g^{(s)}_{1}(z)]= 0,\;\partial_z[h^{(s)}(z)]= 0,\;g^{(s)}\leq 0,
\label{g-f-s-diagonal}
\end{equation}
leading to an uniform energy density in the static zone,
which is invariant under time translation. For this case we have:
\begin{eqnarray}
{\cal T}^{(s)\;\mbox{\footnotesize DD}}={\cal T}^{(s)\;\mbox{\footnotesize NN}},
\;{\cal T}^{(s)\;\mbox{\footnotesize DN}}={\cal T}^{(s)\;\mbox{\footnotesize ND}}.
\label{T-s}
\end{eqnarray}
This extends the equalities ${\cal T}_{\;\mbox{\footnotesize cas}}^{\;\mbox{\footnotesize DD}}={\cal T}_{\;\mbox{\footnotesize cas}}^{\;\mbox{\footnotesize NN}}$ and 
${\cal T}_{\;\mbox{\footnotesize cas}}^{\;\mbox{\footnotesize DD}}={\cal T}_{\;\mbox{\footnotesize cas}}^{\;\mbox{\footnotesize NN}}$,
showed in Eqs. (\ref{T-cas}),
from the vacuum to any other diagonal field state.

For initial field states with 
the density matrix having nonzero off-diagonal elements, 
the function $g^{(s)}$ can have spacetime dependence,
what leads to an energy density non-invariant under time translation. 
For two cavities with different boundary conditions, but the same length $L_0$ and 
values for $\left\langle \hat{a}_{n}^{\dag }\hat{a}_{n^{\prime }}\right\rangle$ we have:
\begin{eqnarray}
{\cal T}_{\left\langle \hat{a}^{\dag}\hat{a}\right\rangle}^{(s)\;\mbox{\footnotesize DD}}
={\cal T}_{\left\langle \hat{a}^{\dag}\hat{a}\right\rangle}^{(s)\;\mbox{\footnotesize NN}},
\;{\cal T}_{\left\langle \hat{a}^{\dag}\hat{a}\right\rangle}^{(s)\;\mbox{\footnotesize DN}}
={\cal T}_{\left\langle \hat{a}^{\dag}\hat{a}\right\rangle}^{(s)\;\mbox{\footnotesize ND}}.
\label{T-s-a-adaga-a}
\end{eqnarray}
On the other hand, for the same values of
$\left\langle \hat{a}_{n}\hat{a}_{n^{\prime }}\right\rangle$, 
we get:
\begin{eqnarray}
{\cal T}_{\left\langle \hat{a}\hat{a}\right\rangle}^{(s)\;\mbox{\footnotesize DD}}
=-{\cal T}_{\left\langle \hat{a}\hat{a}\right\rangle}^{(s)\;\mbox{\footnotesize NN}},
\;{\cal T}_{\left\langle \hat{a}\hat{a}\right\rangle}^{(s)\;\mbox{\footnotesize DN}}
=-{\cal T}_{\left\langle \hat{a}\hat{a}\right\rangle}^{(s)\;\mbox{\footnotesize ND}}.
\label{T-s-a-a}
\end{eqnarray}
However, different and suitable choices for $\left\langle \hat{a}_{n}\hat{a}_{n^{\prime }}\right\rangle$,
for instance $\left\langle \hat{a}_{n}\hat{a}_{n^{\prime }}\right\rangle_1$ and $\left\langle \hat{a}_{n}\hat{a}_{n^{\prime }}\right\rangle_2$
can produce: 
\begin{eqnarray}
{\cal T}_{\left\langle \hat{a}\hat{a}\right\rangle_1}^{(s)\;\mbox{\footnotesize DD}}
={\cal T}_{\left\langle \hat{a}\hat{a}\right\rangle_2}^{(s)\;\mbox{\footnotesize NN}},
\;{\cal T}_{\left\langle \hat{a}\hat{a}\right\rangle_1}^{(s)\;\mbox{\footnotesize DN}}
={\cal T}_{\left\langle \hat{a}\hat{a}\right\rangle_2}^{(s)\;\mbox{\footnotesize ND}},
\label{T-s-a-a-special}
\end{eqnarray}
as showed in Sec. \ref{non-diagonal}, Eq. (\ref{T-zeta-static}).
\section{Non-static situation}
\label{non-static-situation}
\begin{figure}
\epsfig{file=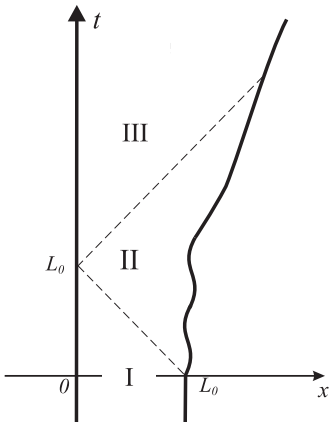,angle=00,width=0.5 \linewidth,clip=}
\caption{Boundary trajectories (solid lines). The dashed lines are null-lines separating the regions
I from II, and II from III.}
\label{static-zones}
\end{figure}
Now, let us examine the cavity in the non-static situation ($t> 0$).
The field modes in Eq. (\ref{field-solution-1}) are formed by left and
right-propagating parts. As causality requires, the field 
in the region I ($v\leq L_0$) (see Fig. \ref{static-zones}) is not
affected by the boundary motion, so that, in this sense, this region is considered as a ``static zone''.
In the region II ($v> L_0$ and $u\leq L_0$), the right-propagating parts 
of the field modes remain unaffected by the boundary motion, so that 
the region II is also a static zone for these modes. 
On the other hand, the left-propagating parts 
in the region II are, in general, affected by the boundary movement. 
In the region III ($u>L_0$), both the left and right-propagating parts are affected. 
In summary, the functions corresponding to the left and right-propagating parts of
the field modes are considered in the static zone if their argument $z$ ($z$ symbolizing $v$ or $u$) is such that $z\leq L_0$. 
We remark that Eqs. (\ref{R-static}), (\ref{f-static}), (\ref{g1-static}) and (\ref{g2-static}) are valid not only in the static situation ($t\leq 0$) but also in the static zone ($t\pm x \leq L_0$).

For a certain spacetime point $(\tilde{t},\tilde{x})$, 
the field operator $\hat{\psi}(\tilde{t},\tilde{x})$ is known if its left and right-propagating
parts, taken over, respectively, the null lines
$v=z_1$ and $u=z_2$ (where $z_1=\tilde{t}+\tilde{x}$ and $z_2=\tilde{t}-\tilde{x}$), are known; or, in other words,
$\hat{\psi}(\tilde{t},\tilde{x})$ is known if $R(v)\vert_{v=z_1}$ and $R(u)\vert_{u=z_2}$ are known.
Cole and Schieve \cite{Cole-Schieve-1995} proposed a recursive 
method to obtain the function $R$ for a general law of motion
of the boundary, tracing back a sequence of null lines until a null line gets into the static zone
where the $R$ function is known. For a brief discussion of this method,
let us assume that $(\tilde{t},\tilde{x})$ belongs to the region III, and that 
the null line $v=z_1$ intersects the moving mirror trajectory
at the point $[t_1,L(t_1)]$ (see Ref. \cite{Cole-Schieve-1995}). We have $R(v)\vert_{v=z_1}=R[t_1+L(t_1)]$. Using the Moore 
equation (\ref{Moore-equation}), we get $R(v)\vert_{v=z_1}=R(u)\vert_{u=t_1-L(t_1)}+2$, so that, as pointed by
Cole and Schieve, ``the value of $R$ increases by 2 every time there is a reflection off the
moving wall" \cite{Cole-Schieve-1995}. 
If $t_1-L(t_1)\leq L_0$, then the null line $u=t_1-L(t_1)$ is already in the static zone,
so that we can write $R(u)\vert_{u=t_1-L(t_1)}=[t_1-L(t_1)]/L_0$, and also
$R(v)\vert_{v=z_1}=[t_1-L(t_1)]/L_0+2$. On the other hand, if $t_1-L(t_1)>L_0$, 
we can draw another null line $v=t_1-L(t_1)$ intersecting the world line 
of the static boundary at the point $[t_1-L(t_1),0]$. Then we get
$R(u)\vert_{u=t_1-L(t_1)}=R(v)\vert_{v=t_1-L(t_1)}$, and 
can write $R(v)\vert_{v=z_1}=R(v)\vert_{v=t_1-L(t_1)}+2$. Assuming that the null line $v=t_1-L(t_1)$
intersects the world line of the moving boundary at the point $[t_2,L(t_2)]$, we get $t_2+L(t_2)=t_1-L(t_1)$,
and we can write: $R(v)\vert_{v=z_1}=R(v)\vert_{v=t_2+L(t_2)}+2$. In the same way, considering
the Moore equation, we can write $R(v)\vert_{v=z_1}=R(u)\vert_{u=t_2-L(t_2)}+4$, and so on,
up to the null line considered is in the static zone,
where the function $R$ is known. In summary, we can write:
$R(\tilde{z})=2n+[\tilde{z}-2\sum_{i=1}^{n}L(t_i)]/L_0$ \cite{Cole-Schieve-1995},
where $n$ is the number of reflections off the moving
boundary, necessary to connect the null line $v=\tilde{z}$ (or $u=\tilde{z}$) to a null line in the static zone.

It is interesting to notice the behavior of the left and right-propagating parts of the field modes in
the cavity. Taking into account the above analysis for the $R$ behavior, for DD and NN cases ($\beta=0$) we get:
$\varphi_{n}^{(0)}(v)\vert_{v=z_1}$$=\varphi_{n}^{(0)}(u)\vert_{u=t_1-L(t_1)}$,
$\varphi_{n}^{(0)}(u)\vert_{u=t_1-L(t_1)}$=$\varphi_{n}^{(0)}(v)\vert_{v=t_1-L(t_1)}$,
and $\varphi_{n}^{(0)}(v)\vert_{v=t_1-L(t_1)}$=$\varphi_{n}^{(0)}(u)\vert_{u=t_2-L(t_2)}$. 
We can see that the value of $\varphi_{n}^{(0)}$ does not change under a reflection off the
static or moving boundaries. The relevant fact is the sequence of the spacetime points 
where the reflections occur, because it will map an initial null line in a non-static zone,
into a certain null line in the static zone, where the value of $\varphi_{n}^{(0)}$ is
determined.
This last comment is also valid for mixed boundary conditions ($\beta=1/2$), but, in this case, we 
observe that the value of $\varphi_{n}^{(1/2)}$ does not change under a reflection off the
static boundary, but changes sign under a reflection off the moving boundary. 

The behavior of the energy density in a cavity, for the vacuum as the initial state,
is described by the function $f$, which obeys the following equation
proposed by Cole and Schieve \cite{Cole-Schieve-2001}:
\begin{eqnarray}
f\left[t+L\left( t\right) \right] &=&f\left[t-L\left(
t\right) \right]{\cal{A}}(t)+{\cal{B}}(t),
\label{f}
\end{eqnarray}
where
\begin{equation}
{\cal{A}}(t)=\left[ \frac{1-L^{\prime }\left(
t\right) }{1+L^{\prime }\left( t\right) }\right]^{2},
\label{f-A}
\end{equation}
\begin{eqnarray}
{\cal{B}}(t)&=&-\frac{1}{12\pi }\frac{L^{\prime \prime \prime }\left(
t\right) }{\left[ 1+L^{\prime }\left( t\right) \right]^{3}\left[
1-L^{\prime }\left( t\right) \right] }\nonumber\\
&&-\frac{1}{4\pi }\frac{L^{\prime \prime 2}\left( t\right) L^{\prime }\left(
t\right) }{\left[ 1+L^{\prime }\left( t\right) \right]^{4}\left[
1-L^{\prime }\left( t\right) \right]^{2}},
\label{f-B}
\end{eqnarray}
which enable us to obtain $f(z)$, and therefore ${\cal T}_{\;\mbox{\footnotesize vac}}$, recursively
in terms of its value $f=f^{({s})}$ in the static zone (see Refs. \cite{Cole-Schieve-1995,Cole-Schieve-2001}).

For an arbitrary initial field state, the behavior of the energy density
in a cavity is described by the function $f$ (related to the vacuum part)
and a new function $g$ (related to the non-vacuum part) proposed in the present paper. 
For the latter one, from Eqs. (\ref{Moore-equation}), (\ref{g1}) and (\ref{g2}) we obtain the following equation:
\begin{equation}
g\left[t+L\left( t \right)\right]=g\left[t-L\left( t \right)\right]{\cal{A}}(t),
\label{g-recorrencia}
\end{equation}
which enable us to get $g(z)$, and therefore ${\cal T}_{\;\mbox{\footnotesize non-vac}}$, recursively
in terms of its value $g(z)=g^{({s})}(z)$ in the static zone.

Now, for a complete analysis involving vacuum and non-vacuum parts,
let us write the equation for the function $h$
(Eq. (\ref{T-em-termos-h})) as follows:
\begin{eqnarray}
h\left[t+L\left( t\right) \right] &=&h\left[t-L\left(
t\right) \right]{\cal{A}}(t)+{\cal{B}}(t).
\label{h}
\end{eqnarray}
The procedure to find $h(z)$, solving recursively the Eq. (\ref{h}),
starts by setting $z=t_1+L\left(t_1\right)$ and tracing the null line $v=t_1+L\left(t_1\right)$.
From Eq. (\ref{h}), after the first reflection traced back on the moving boundary, we get
the relation $h(z)=h\left[t_1-L\left(t_1\right)\right]{\cal{A}}(t_1)+{\cal{B}}(t_1)$. Now,
as explained above, we set
$t_1-L\left(t_1\right)=t_2+L\left(t_2\right)$, and, after
the second reflection on the moving boundary, we get:
\begin{eqnarray}
h(z)&=&h\left[t_2-L\left(
t_2\right) \right]{\cal{A}}(t_2){\cal{A}}(t_1)
+{\cal{B}}(t_2){\cal{A}}(t_1)+{\cal{B}}(t_1)\nonumber.
\end{eqnarray}
This process goes on, up to a null line gets into a static zone, where 
$h=h^{(s)}=f^{(s)}+g^{(s)}$ is known. 
At the end, we get:
\begin{eqnarray}
h(z)&=&h^{(s)}(z){\cal{\tilde{A}}}(z)+{\cal{\tilde{B}}}(z),
\label{h-final}
\end{eqnarray}
where:
\begin{equation}
h^{(s)}(z)=f^{(s)}+g^{(s)}[\tilde{z}(z)],
\label{h-s}
\end{equation}
\begin{equation}
{\cal{\tilde{A}}}(z)=\Pi_{i=1}^{n}{\cal{A}}(t_{i}),
\end{equation}
\begin{eqnarray}
{\cal{\tilde{B}}}(z)&=&{\cal{B}}(t_{n})\Pi
_{i=1}^{n-1}{\cal{A}}(t_{i})+{\cal{B}}(t_{n-1})\Pi
_{i=1}^{n-2}{\cal{A}}(t_{i})\nonumber\\
&&+...+{\cal{B}}(t_{2}){\cal{A}}(t_{1})+{\cal{B}}(t_{1}).
\end{eqnarray}
The number $n$ of reflections and the sequence of instants $t_1,...t_n$ depend on 
$z$. In Eq. (\ref{h-s}), we have $\tilde{z}(z)=t_n-L(t_n)$. 
From this process, we observe that the value of $f$ does not change under a reflection off the
static boundary, but it changes under a reflection off the moving boundary, as expected.
We also note, from Eq. (\ref{h}), that the function $h$, after a reflection,
changes in the same way for all boundary conditions considered in the present work
(for the same boundary motion). 
We remark that, in Eq. (\ref{h-final}), the functions ${\cal\tilde{A}}$ and ${\cal\tilde{B}}$ depend only on the 
law of motion of the moving mirror, whereas all the dependence on the boundary conditions considered in the present paper,
and on the initial field state, are stored in the static region function $h^{(s)}(z)$.
We also observe that, for a generic law of motion, 
${\cal{\tilde{A}}}$ and ${\cal{\tilde{B}}}$ are different functions,
with the following properties: 
\begin{equation}
{\cal{\tilde{A}}}(z)>0\;\forall\;z,\;
{{\cal{\tilde{A}}}(z< L_0)}=1,\;
{\cal{\tilde{B}}}(z< L_0)=0,
\label{A-b-cond}
\end{equation}
which are directly obtained from Eqs. 
(\ref{f-A}) and (\ref{f-B}).

The energy density $\cal{T}$ can be given as in Eq. (\ref{T}), but now 
with the parts ${\cal T}_{\mbox{\footnotesize vac}}$ and ${\cal T}_{\mbox{\footnotesize non-vac}}$ 
rewritten as: 
\begin{equation}
{\cal T}_{\mbox{\footnotesize vac}}=-f^{(s)}\left[{\cal{\tilde{A}}}(u)+{\cal{\tilde{A}}}(v)\right]
-{\cal{\tilde{B}}}(u)-{\cal{\tilde{B}}}(v),
\label{T-vac-A-B}
\end{equation}
\begin{equation}
{\cal T}_{\mbox{\footnotesize non-vac}}=-g^{(s)}[\tilde{z}(u)]{\cal{\tilde{A}}}(u)
-g^{(s)}[\tilde{z}(v)]{\cal{\tilde{A}}}(v).
\label{T-non-vac-B}
\end{equation}
The instantaneous force acting on the moving boundary, in this two-dimensional model, 
is given by ${\cal T}(t,L(t))$.
From Eqs. (\ref{T-em-termos-h}), (\ref{T-vac-A-B}) and (\ref{T-non-vac-B}) we see that two cavities
with different boundary conditions and the same initial length $L_0$,
presenting different vacuum and non-vacuum contributions for their initial energy densities,
but with the same total energy density ${\cal T}^{(s)}$
in the static zone, for a same (but arbitrary) law of motion 
will always exhibit identical evolutions in time for their energy
densities ${\cal T}$. 

\subsection{Diagonal states}
\label{diagonal}

Now, let us analyze and compare the behavior of 
${\cal T}_{\mbox{\footnotesize vac}}$, ${\cal T}_{\mbox{\footnotesize non-vac}}$ and
${\cal T}$ for the initial field states for which the density matrix is diagonal
in the Fock basis. From Eqs. (\ref{g-f-s-diagonal}) and
(\ref{T-non-vac-B}) we obtain:
\begin{equation}
{\cal T}_{\mbox{\footnotesize non-vac}}=-g^{(s)}\left[{\cal{\tilde{A}}}(u)
+{\cal{\tilde{A}}}(v)\right],
\label{T-non-vac-B-diagonal}
\end{equation}
where  $g^{(s)}$ is a constant factor. From Eqs.  (\ref{T-vac-A-B}) and (\ref{T-non-vac-B-diagonal})
we can see that all spacetime dependence 
in ${\cal T}$ comes from ${\cal{\tilde{A}}}$ and ${\cal{\tilde{B}}}$.
For this case we have:
\begin{equation}
{\cal T}^{\;\mbox{\footnotesize DD}}={\cal T}^{\;\mbox{\footnotesize NN}},\;
{\cal T}^{\;\mbox{\footnotesize DN}}={\cal T}^{\;\mbox{\footnotesize ND}}. 
\label{TDD-TNN-e-TDN-TND}
\end{equation}
Since Eq. (\ref{TDD-TNN-e-TDN-TND}) is valid for any law of motion, 
it generalizes the Eq. (\ref{T-s}), and because it is also valid
for any diagonal state, it extends the conclusion obtained in Ref. \cite{Alves-Granhen-2008},
where formulas correspondent to Eq. (\ref{TDD-TNN-e-TDN-TND}) were obtained just considering
the vacuum as the initial field state.

The total energy stored in the cavity as a function of time 
can be written as $\mathcal{E}(t) = \int_{0}^{L(t)}{\cal T}(t,x)\;dx$.
For diagonal initial states, we have:
\begin{equation}
\mathcal{E}(t)= -h^{(s)}{\cal F}_{1}(t)-{\cal F}_2(t),
\label{E-diagonal}
\end{equation}
where ${\cal F}_{1}(t)=\int_{0}^{L(t)}\left[{\cal{\tilde{A}}}(u)+{\cal{\tilde{A}}}(v)\right]dx$
and ${\cal F}_{2}(t)=\int_{0}^{L(t)}\left[{\cal{\tilde{B}}}(u)+{\cal{\tilde{B}}}(v)\right]dx$.
We emphasize that  ${\cal F}_{1}$ and ${\cal F}_{2}$ depend only on the law of motion of the boundary,
whereas all dependence on the initial field state or type of boundary condition is stored
in the coefficient $h^{(s)}$. 

Since the functions ${\cal{\tilde{A}}}$ and ${\cal{\tilde{B}}}$ are, in general, different one from the other,
we conclude that ${\cal T}_{\mbox{\footnotesize vac}}$, ${\cal T}_{\mbox{\footnotesize non-vac}}$ and
${\cal T}$ exhibit, in general, different structures. On the other hand, one could ask about the conditions under
which ${\cal T}_{\mbox{\footnotesize vac}}$, ${\cal T}_{\mbox{\footnotesize non-vac}}$ and
${\cal T}$ would exhibit the same structure (the same curve apart from a positive scale factor or an additive constant) 
so that, for instance, if there are peaks and
valleys they would be at the same positions in the cavity.
One condition is provided by laws of motion for which
${\cal{\tilde{B}}}(z)$ and ${\cal{\tilde{A}}}(z)$ have a linear relation as given by:
\begin{equation}
{\cal{\tilde{B}}}(z)= k_1{\cal{\tilde{A}}}(z)+k_2,
\label{B-A-relation}
\end{equation}
where $k_1$ e $k_2$ are constants. From the properties in Eq. (\ref{A-b-cond}), we get $k_1=-k_2$, resulting:
\begin{equation}
{\cal{\tilde{B}}}(z)=k_1[{\cal{\tilde{A}}}(z)-1].
\label{B-approx-A}
\end{equation}
From (\ref{T-vac-A-B}), (\ref{T-non-vac-B}) and (\ref{B-approx-A}), we get:
\begin{equation}
{\cal T}_{\mbox{\footnotesize vac}}=-(f^{(s)}+k_1)[{\cal{\tilde{A}}}(u)+{\cal{\tilde{A}}}(v)]+2k_1,
\label{Pe}
\end{equation}
\begin{equation}
{\cal T}_{\mbox{\footnotesize non-vac}}=-g^{(s)}[{\cal{\tilde{A}}}(u)+{\cal{\tilde{A}}}(v)],
\label{De}
\end{equation}
\begin{equation}
{\cal T}=-(f^{(s)}+g^{(s)}+k_1)[{\cal{\tilde{A}}}(u)+{\cal{\tilde{A}}}(v)]+2k_1.
\label{Moleque}
\end{equation}
In the energy densities showed in Eqs. (\ref{Pe})-(\ref{Moleque}),
if the constant factors multiplying ${\cal{\tilde{A}}}(u)+{\cal{\tilde{A}}}(v)$ are different from zero and
have the same sign, 
as the ratio $\sigma(z)={\cal{\tilde{B}}}(z)/[{\cal{\tilde{A}}}(z)-1]$ 
becomes more approximately equal to a constant value $k_1$, more
the structures of ${\cal T}_{\mbox{\footnotesize vac}}$, ${\cal T}_{\mbox{\footnotesize non-vac}}$ and
${\cal T}$ become similar one to the other. 
Next, we exemplify these situations in
the context of the vacuum and thermal field, which are examples of a diagonal initial field  states. 

\begin{figure}
\epsfig{file=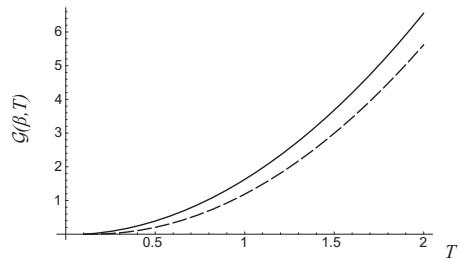,angle=00,width=0.7 \linewidth,clip=}
\caption{Plot of ${\cal G}(\beta,T)$. The solid curve corresponds to ${\cal G}(1/2,T)$, whereas the dashed one 
to ${\cal G}(0,T)$.}
\label{coefficient-thermal}
\end{figure}

For the thermal state we take into account that 
$\langle \hat{{a}}^{\dagger}_{n}\hat{a}_{n'}\rangle=\delta_{nn'}\overline{n}(n,\beta)$
and $\langle \hat{a}_{n}\hat{a}_{n^\prime}\rangle=\langle \hat{{a}}^{\dagger}_{n}\hat{a}^{\dagger}_{n'}\rangle=0$, where 
$
\overline{n}\left(n,\beta\right)={1}/({e^{(n+\beta) /T}-1}),
$
with the Boltzmann constant ($k_B$) equal to 1.
For this initial state, the energy density ${\cal T}_{\;\mbox{\footnotesize non-vac}}$, now relabeled as ${\cal T}_{T}$, is given by:
\begin{eqnarray}
{\cal T}_{T}&=&\frac{\pi{\left|\gamma\right|}^{2}}{2L_0^2}{\cal G}(\beta,T)\left[{\cal{\tilde{A}}}(u)+{\cal{\tilde{A}}}(v)\right],
\label{termico}
\end{eqnarray}
where ${\cal G}(\beta,T)=\sum^{\infty }_{n=1-2\beta}\left[({n+\beta})/({e^{(n+\beta) /T}-1})\right]$
plays the role of a scale factor. Since ${\cal G}(1/2,T)>{\cal G}(0,T)$ (see Fig. \ref{coefficient-thermal}),
for $T>0$ we have:
\begin{equation}
{{\cal T}_{T}}^{\;\mbox{\footnotesize mixed}}>{{\cal T}_{T}}^{\;\mbox{\footnotesize non-mixed}},
\end{equation}
where the superscript ``mixed'' means ND or DN cases, whereas ``non-mixed'' means NN or DD cases.

Note that if we set up two different cavities, one of them with non-mixed and
the other one with mixed boundary conditions, both with a thermal field as the initial state
with temperatures given, respectively, by $T_1$ and $T_2$ such that
$
{\cal G}(1/2,T_2)-{\cal G}(0,T_1)=-1/16,
$
we get for both cases the same initial configurations of energy densities
in the static zone. Then for a same arbitrary law of motion for both
cavities, they will exhibit the same time evolutions for the energy densities ${\cal T}$
(and hence the same force acting on the boundaries). An interesting case occurs if we set up $T_1\approx0.34$ and $T_2=0$. In this case we obtain
that the static and dynamical Casimir effect for the vacuum state in the mixed case (where
the Casimir static force is repulsive) is mimicked by the 
non-mixed case with the mentioned specific temperature. 

\begin{figure}
\epsfig{file=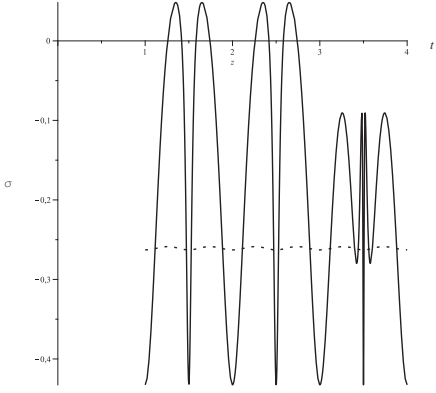,angle=00,width=0.7 \linewidth,clip=}
\caption{The ratio $\sigma(z)={\cal{\tilde{B}}}(z)/[{\cal{\tilde{A}}}(z)-1]$
for the law of motion given in Eq. (\ref{eq-movimento-andreata-dodonov}). 
The dashed line corresponds to $\epsilon=0.01$, wheres the solid line
corresponds to $\epsilon=0.1$.}
\label{approximation}
\end{figure}

\begin{figure}
\epsfig{file=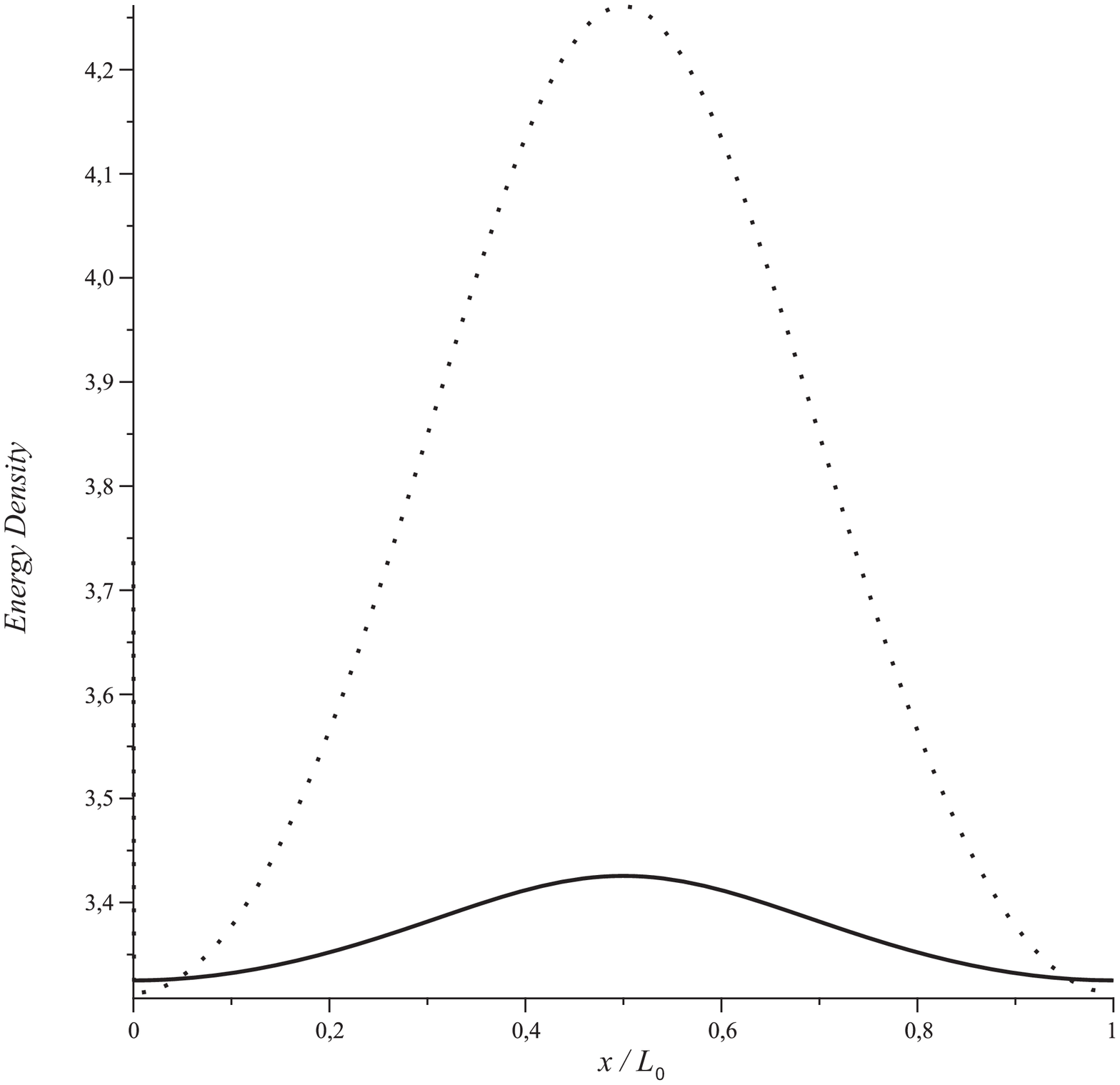,angle=00,width=0.7 \linewidth,clip=}
\caption{The energy densities ${\cal T}_{\mbox{\footnotesize vac}}$ (solid line)
and ${\cal T}_{\mbox{\footnotesize non-vac}}$ (dotted line) at the instant
$t=1$, for both DD and NN cases. The law of motion considered is
showed in Eq. (\ref{eq-movimento-andreata-dodonov}), with the parameter $\epsilon=0.01$.
For the curve correspondent to ${\cal T}_{\mbox{\footnotesize non-vac}}$,
it is being considered a the thermal bath with temperature $T=1$ as the initial field state.}
\label{epsilon-001}
\end{figure}

\begin{figure}
\epsfig{file=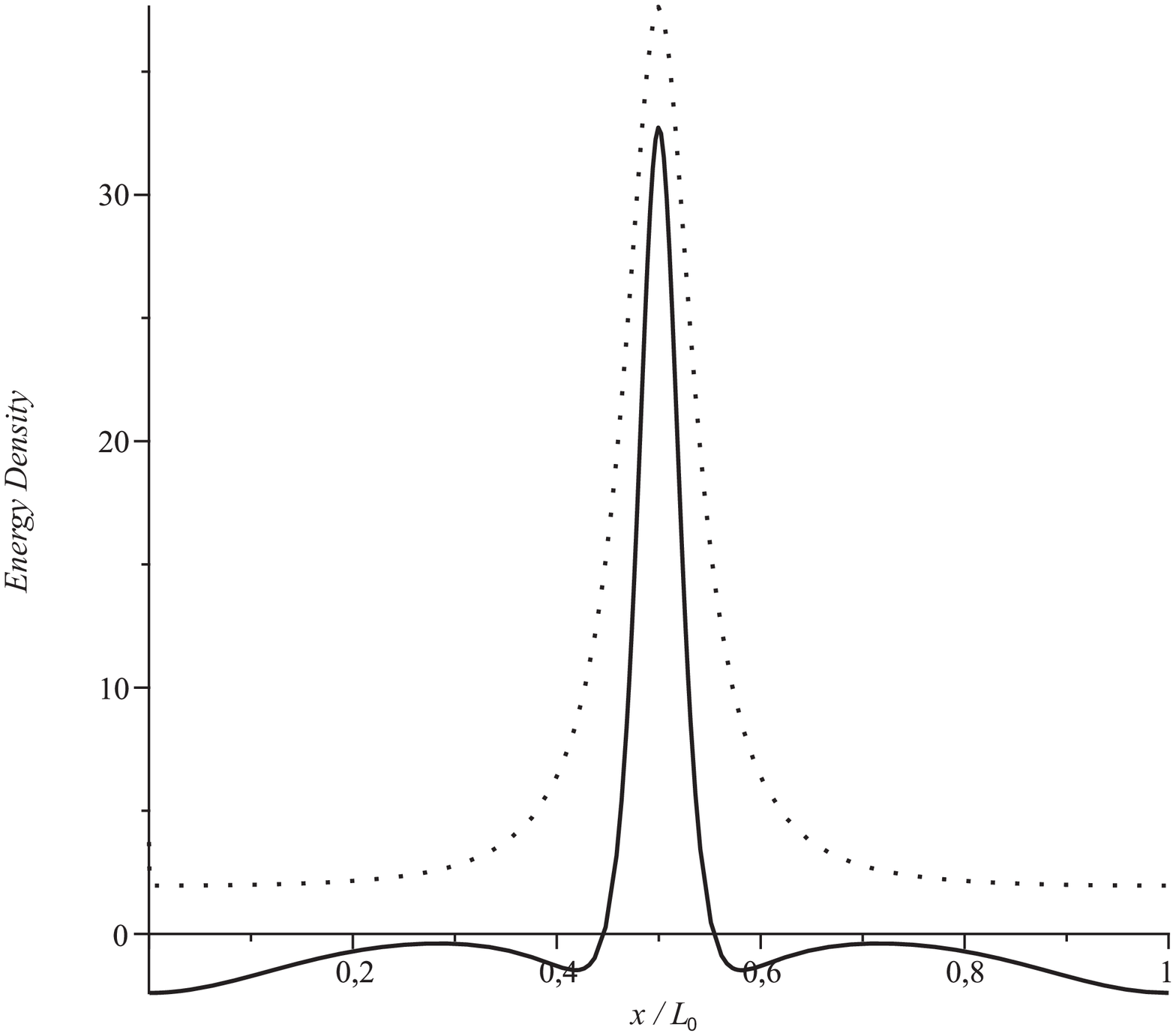,angle=00,width=0.7 \linewidth,clip=}
\caption{The energy densities ${\cal T}_{\mbox{\footnotesize vac}}$ (solid line)
and ${\cal T}_{\mbox{\footnotesize non-vac}}$ (dotted line) at the instant
$t=1$, for both DD and NN cases. The law of motion considered is
showed in Eq. (\ref{eq-movimento-andreata-dodonov}), with the parameter $\epsilon=0.1$.
For the curve correspondent to ${\cal T}_{\mbox{\footnotesize non-vac}}$,
it is being considered a the thermal bath with temperature $T=1$ as the initial field state.}
\label{epsilon-01}
\end{figure}

Let us now investigate a particular movement described by:
\begin{equation}
L\left( t\right) =L_{0}\left[1+\varepsilon \sin \left( \frac{p\pi t}{L_{0}}%
\right)\right],
\label{eq-movimento-andreata-dodonov}
\end{equation}
where $L_0=1$, $p=2$ and $\epsilon$ is a dimensionless parameter.
This oscillatory boundary motion is
investigated in several papers (see, for instance, Refs. \cite{Dodonov-JMP-1993,estados-coerentes-2}).
In these papers the calculations
were developed in the context of analytical approximate methods, considering small amplitudes
of oscillation ($|\epsilon| \ll 1$).
Here, we investigate the law of
motion in Eq. (\ref{eq-movimento-andreata-dodonov}) for 
$\epsilon=0.01$ and $\epsilon=0.1$. The former value
is in better agreement with the condition $|\epsilon| \ll 1$
than the latter one. Our intention now
is to verify, from an exact point of view, the similarity
between the structures of ${\cal T}_{\mbox{\footnotesize vac}}$, ${\cal T}_{\mbox{\footnotesize non-vac}}$ and
${\cal T}$ for both values of $\epsilon$.
In Fig. \ref{approximation} we plot the ratio $\sigma$ 
for the two values: $\epsilon=0.01$ and $\epsilon=0.1$. We observe that the ratio $\sigma$ is
more approximately constant for $\epsilon=0.01$ (dashed line) than for
$\epsilon=0.1$ (solid line), so that,
since $f^{(s)}+k_1<0$ and $g^{(s)}\leq0$,
we expect that ${\cal T}_{\mbox{\footnotesize vac}}$, ${\cal T}_{\mbox{\footnotesize non-vac}}$ and
${\cal T}$ exhibit more similarity for the former than for the latter value of $\epsilon$,
for any diagonal initial state or any of the boundary conditions considered in the present paper.
In fact, this is what can be visualized in Figs. \ref{epsilon-001} and \ref{epsilon-01},
where we show ${\cal T}_{\mbox{\footnotesize vac}}$ (dotted lines) and ${\cal T}_{\mbox{\footnotesize non-vac}}$
(solid lines) for the law of motion (\ref{eq-movimento-andreata-dodonov}) and the thermal bath with 
temperature $T=1$ as the diagonal initial field state.
In Fig. \ref{epsilon-001}, for the case $\epsilon=0.01$, both dotted and solid lines have
their minimum and maximum values at the same positions, as predicted
by the approximate analytical approach considered by
Andreata and Dodonov \cite{estados-coerentes-2}, which was based on $\epsilon\ll 1$. 
On the other hand, in Fig. \ref{epsilon-01} (case $\epsilon=0.1$), if compared to the dotted line,
the solid line has additional maximum points at $x=0.28$ and $x=0.72$,
and also additional minimum points at $x=0.42$ and $x=0.58$, so that
differences between the structures of ${\cal T}_{\mbox{\footnotesize vac}}$ and ${\cal T}_{\mbox{\footnotesize non-vac}}$
start to become more evident. 
%
\subsection{Non-diagonal states}
\label{non-diagonal}
When we consider non-diagonal states, the function $g^{(s)}$ 
in Eq. (\ref{T-non-vac-B}) can depend on $\tilde{z}(z)$ 
or, in other words, it can depend on 
the null line in the static zone where it is being calculated.
Since Eq. (\ref{g2-static}) depends on $\gamma^2$, now ${\cal T}$ can be different
for DD and NN cases (it is the same for DN and ND cases). We can see that ${\cal T}_{\mbox{\footnotesize vac}}$, ${\cal T}_{\mbox{\footnotesize non-vac}}$ and
${\cal T}$ exhibit, in general, different structures. 

The total energy stored in the cavity as a function of time 
can be written as:
\begin{equation}
\mathcal{E}(t)= -f^{(s)}{\cal F}_{1}(t)-{\cal F}_2(t)-{\cal F}_3(t),
\label{E-non-diagonal}
\end{equation}
where ${\cal F}_{3}(t)=\int_{0}^{L(t)}\left[g^{(s)}[\tilde{z}(u)]{\cal{\tilde{A}}}(u)
+g^{(s)}[\tilde{z}(v)]{\cal{\tilde{A}}}(v)\right]dx$.
The function ${\cal F}_{3}$ in Eq. (\ref{E-non-diagonal}) depends on the type of
boundary condition, the initial field state and the law of motion considered.
Next, we exemplify these situations in
the context of a coherent and a Schrödinger cat state, which are examples of non-diagonal initial field  states. 

The coherent state is defined as an eigenstate of the annihilation operator: 
$\hat{a}_{n}\left|\alpha\right\rangle=\alpha\delta_{nn_{0}}\left|\alpha\right\rangle$, where 
$\alpha={\left|\alpha\right|}e^{i\theta}$ and
$n_{0}$ is related to the frequency of the excited mode \cite{Glauber}. 
The Schrödinger cat state \cite{Gato,Jeong-2002} is defined 
as a superposition of coherent states $\left|\Psi\right\rangle=N\left(\left|\alpha\right\rangle+e^{i\phi}\left|-\alpha\right\rangle\right)$, 
where $\left|-\alpha\right\rangle$ has the same amplitude than $\left|\alpha\right\rangle$ but with
a phase shift of $\pi$, and the normalization constant $N$ is given by $N=(2+2e^{-2{\left|\alpha\right|}^{2}}\cos{\phi})^{-{1}/{2}}$.
We combine into a single formula the results for the energy density ${\cal T}_{\;\mbox{\footnotesize non-vac}}$
(now relabeled as ${\cal T}_{\zeta}$) for both coherent and cat states, in the following manner:
\begin{eqnarray}
{\cal T}_{\zeta}&=&{\cal T}_{\left\langle \hat{a}^{\dag}\hat{a}\right\rangle_{\zeta}}+{\cal T}_{\left\langle \hat{a}\hat{a}\right\rangle_{\zeta}},
\label{energia-coerente-gato}
\end{eqnarray}
where
\begin{eqnarray}
{\cal T}_{\left\langle \hat{a}^{\dag}\hat{a}\right\rangle_{\zeta}}&=&g^{(s)}_1\left[{\cal{\tilde{A}}}(u)+{\cal{\tilde{A}}}(v)\right],
\label{T-a-adaga-a}
\\
{\cal T}_{\left\langle \hat{a}\hat{a}\right\rangle_{\zeta}}&=&g^{(s)}_2[\tilde{z}(u)]{\cal{\tilde{A}}}(u)
+g^{(s)}_2[\tilde{z}(v)]{\cal{\tilde{A}}}(v),
\label{T-a-a}
\end{eqnarray}
with
\begin{equation}
g^{(s)}_1=\frac{{\left|\gamma\right|}^{2}{\left|\alpha\right|}^{2} \pi}{2L_0^2}(n_0+\beta){\cal C}(\phi,|\alpha|)^{\eta},
\label{gs1-coerente}
\end{equation}
and
\begin{equation}
g^{(s)}_2(z)=-\frac{{\left|\alpha\right|}^{2}{\gamma}^{2}\pi}{2L_0^2}(n_0+\beta)\cos{[2\pi(n_0+\beta)z/L_0-2\theta]},
\label{gs2-equation}
\end{equation}
where ${\cal C}(\phi,|\alpha|)=(1-e^{-2{\left|\alpha \right|}^{2}} \cos{\phi}) / (1+e^{-2{\left|\alpha \right|}^{2}} \cos{\phi})$.
%
%
%
%
%
For $\eta=0$ we recover the coherent case, whereas for $\eta=1$ the cat state energy density is obtained. 

\begin{figure}
\epsfig{file=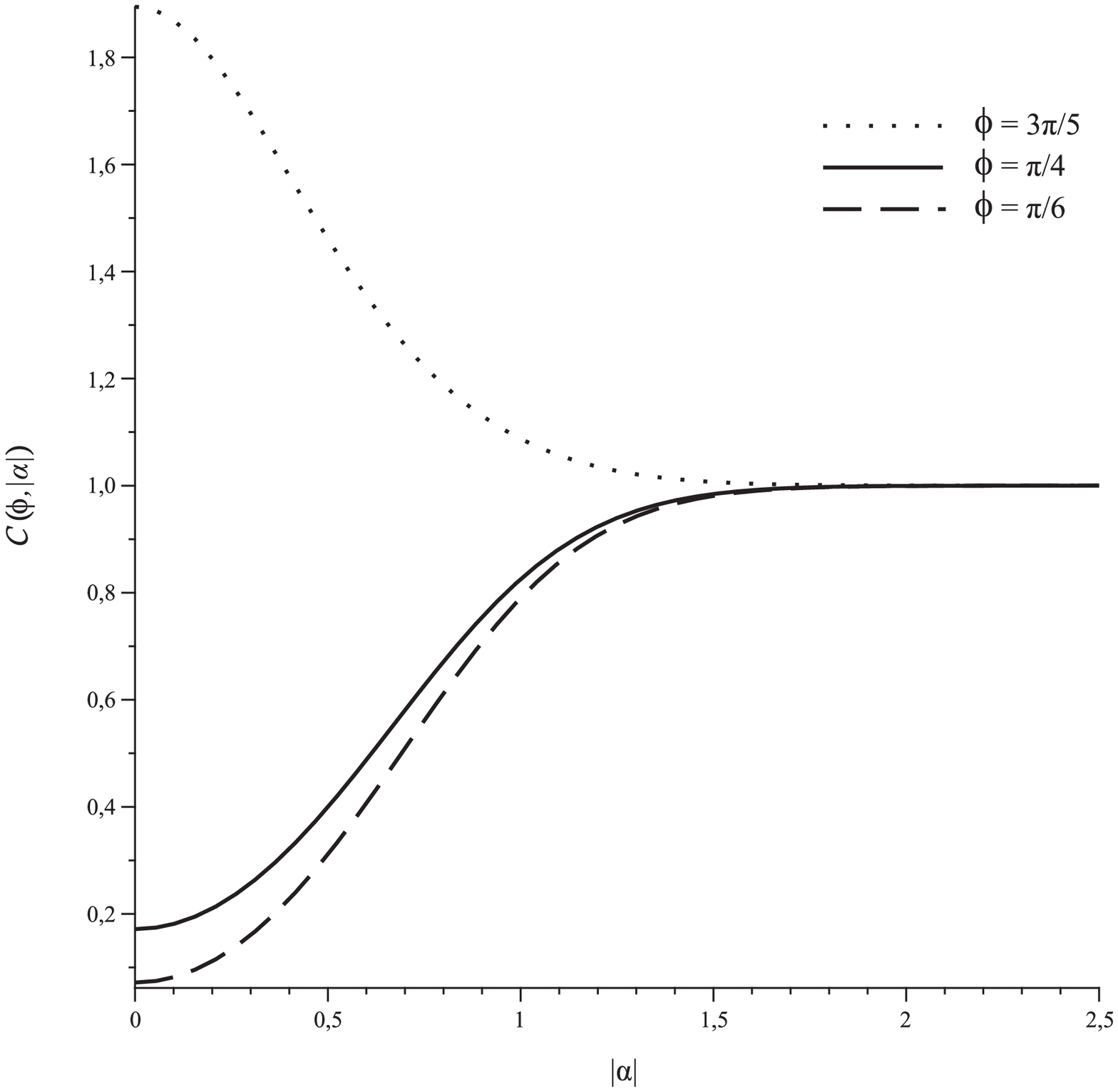,angle=00,width=0.6 \linewidth,clip=}
\caption{${\cal C}(\phi,|\alpha|)$ as function of $\left|\alpha \right|$, for different values of $\phi$.}
\label{coefficient-alpha}
\end{figure}

The two coherent states $|\alpha\rangle$ and $|-\alpha\rangle$ are not orthogonal to each other 
and their overlap $|\langle\alpha|-\alpha\rangle|^2=e^{-4|\alpha|^2}$ decreases exponentially with
$|\alpha|$. For $|\alpha|$=2 the overlap is approximately zero \cite{Jeong-2005}, then we expect that the cat state behaves like two coherent states. From Eq. (\ref{energia-coerente-gato}) for $\left|\alpha \right|\approx2$ we have ${\cal C}\approx 1$ (see Fig. \ref{coefficient-alpha}),
hence the energy density for cat states and coherent states become the same.  Both these energy densities
are also the same, now for any value of $|\alpha|$, if 
we consider the Yurke-Stoler state (a cat state with phase $\phi=\frac{\pi}{2}$).

Observe that $g^{(s)}_1$ is constant in the static zone, whereas $g^{(s)}_2$ is spacetime dependent. From Eq. (\ref{T-non-vac-B-diagonal}) and (\ref{T-a-adaga-a}) we see that ${\cal T}_{\mbox{\footnotesize non-vac}}$ for diagonal states
presents the same structure of ${\cal T}_{\left\langle \hat{a}^{\dag}\hat{a}\right\rangle_{\zeta}}$.
An interesting result occurs when one considers the cat state with $\phi=0$ (odd cat state) and takes the limit $\left|\alpha\right| \rightarrow 0$ 
(leading to a single particle state) in Eq. (\ref{energia-coerente-gato}). In this case we obtain $g^{(s)}=\left(\pi/2L^2_{0}\right)\left(n_{0}+\beta\right)$ and ${\cal T}_{\left\langle \hat{a}\hat{a}\right\rangle_{\zeta}}=0$. This means that the odd cat state in the limit $\left|\alpha\right| \rightarrow 0$ presents its energy density with the same structure of a 
single particle diagonal state as expected. 

\begin{figure}
\epsfig{file=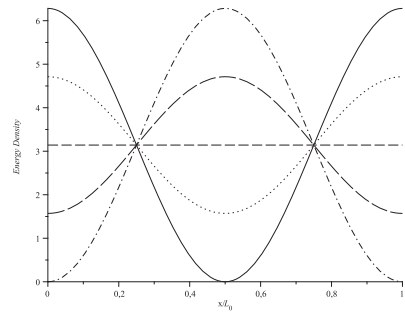,angle=00,width=0.6 \linewidth,clip=}
\caption{The normalized energy density ${\cal T}_{\zeta}^{(s)}\vert_{\eta=0}/|\alpha|^2$ (coherent state) with 
$n_0=1$, as function of $x/L_0$. The solid, dotted, dashed,
dashed-dotted and long-dashed lines can have different meanings, for instance:
(a) In the DD case for $t=0$, they represent, respectively,
the energy density for $\theta = 0$, $\pi/6$, $\pi/4$,
$\pi/2$ and $\pi/3$; 
(b) In the DD case for $\theta=0$, they represent the time evolution of the energy density
in the static situation ($t<0$), corresponding, respectively, to $t = 0$, $-1/6$, $-1/4$,
$-1/2$ and $-1/3$. Also
the correspondent NN situations can be obtained via Eq. (\ref{T-zeta-static}).}
\label{coherent-static}
\end{figure}

From Eq. (\ref{gs2-equation}) we see that $\theta$ contributes to the shape of the energy density in
the static zone. The Fig. \ref{coherent-static} shows the normalized coherent energy density
${\cal T}_{\zeta}^{(s)}\vert_{\eta=0}/|\alpha|^2$ as function of the position in a static situation,
for fixed $\theta$ and several values of time and vice versa. Despite the fact that the energy
density ${\cal T}_{\left\langle \hat{a}\hat{a}\right\rangle_{\zeta}}$ depends on $\theta$,
the total static energy ${\cal E}^{(s)}$
does not depend on this variable, as shown in Eq. (\ref{int-g2}). On the other hand,
since ${\cal{F}}_3(t)$ can depend on $\theta$,
from Eq. (\ref{E-non-diagonal}) we see that the dynamical total energy $\cal E$ can depend on 
this variable.

\begin{figure}
\epsfig{file=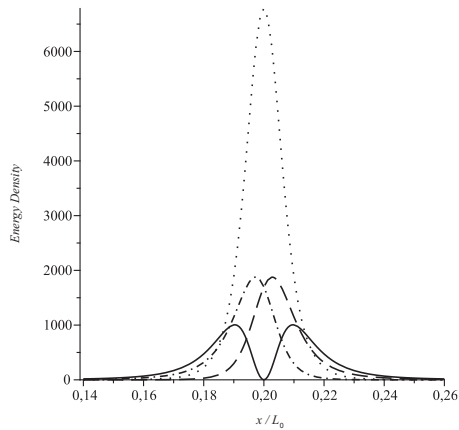,angle=00,width=0.7 \linewidth,clip=}
\caption{The normalized coherent energy density ${{\cal T}_{\zeta}}/|\alpha|^2$ for $n_{0}=1$,
$t=50.3$ and the law of motion given in Eq. (\ref{eq-movimento-andreata-dodonov}) with $\epsilon=0.01$,
as function of $x/L_0$. The solid line corresponds to both: ${{\cal T}_{\zeta}}^{DD}$ with $\theta=0$ and ${{\cal T}_{\zeta}}^{NN}$ with $\theta=\pi/2$. The dotted one corresponds to ${{\cal T}_{\zeta}}^{NN}$ with $\theta=0$ and also ${{\cal T}_{\zeta}}^{DD}$ with $\theta=\pi/2$. The dashed-dotted line  corresponds to ${{\cal T}_{\zeta}}^{DN}$ with $\theta=0$ and also  ${{\cal T}_{\zeta}}^{ND}$ with $\theta=\pi/2$. The dashed line shows ${{\cal T}_{\zeta}}^{ND}$ with $\theta=0$ and also  ${{\cal T}_{\zeta}}^{DN}$ with $\theta=\pi/2$.}
\label{coherent}
\end{figure}

For different boundary conditions, the same configuration of energy density can be obtained by choosing
different values of $\theta$ in each case.
Note that under a phase shift $\theta \rightarrow \theta + \frac{\pi}{2}$ we have
the following symmetries: 
\begin{equation}
g_2^{(s)\;{\mbox{\footnotesize DD}}}\vert_{\theta}=g_2^{(s)\;{\mbox{\footnotesize NN}}}\vert_{\theta + \frac{\pi}{2}},\;
g_2^{(s)\;{\mbox{\footnotesize DN}}}\vert_{\theta}=g_2^{(s)\;{\mbox{\footnotesize ND}}}\vert_{\theta + \frac{\pi}{2}}, 
\end{equation}
thus we get:
\begin{equation}
{\cal T}_\zeta^{(s)\;{\mbox{\footnotesize DD}}}\vert_{\theta}={\cal T}_\zeta^{(s)\;{\mbox{\footnotesize NN}}}\vert_{\theta + \frac{\pi}{2}},\;{\cal T}_\zeta^{(s)\;{\mbox{\footnotesize DN}}}(z)\vert_{\theta}={\cal T}_\zeta^{(s)\;{\mbox{\footnotesize ND}}}\vert_{\theta + \frac{\pi}{2}}
\label{T-zeta-static}
\end{equation}
or, in other words, under the mentioned phase shift the energy densities for DN and ND (and also for DD and NN as
shown in Fig. \ref{coherent-static}) 
in the static zone are the same. Then, from Eqs. (\ref{T-em-termos-h}) and  (\ref{h-final}) we obtain that the energy density at any instant $t$  satisfies: 
\begin{eqnarray}
{\cal T}_{\zeta}^{\;{\mbox{\footnotesize DD}}}\vert_{\theta}
={\cal T}_{\zeta}^{\;{\mbox{\footnotesize NN}}}\vert_{\theta + \frac{\pi}{2}},
\;{\cal T}_{\zeta}^{\;{\mbox{\footnotesize DN}}}\vert_{\theta}
={\cal T}_{\zeta}^{\;{\mbox{\footnotesize ND}}}\vert_{\theta + \frac{\pi}{2}}.
\label{ta-cabando}
\end{eqnarray}
In Fig. \ref{coherent} we show the normalized coherent energy density ${{\cal T}_{\zeta}}/|\alpha|^2$ 
for DD, NN, ND and DN cases, obtained via numerical calculations based on the exact 
Eqs. (\ref{T-a-adaga-a}) and (\ref{T-a-a}).
In this figure, in which $\epsilon=0.01$, the result for the DD case is 
in good agreement with the approximate analytical one obtained by Andreata and Dodonov for this
boundary condition \cite{estados-coerentes-2}.

\section{Final Comments}
\label{final-comments}

In summary, we extended to the case of an arbitrary initial field state
the exact method proposed by Cole and Schieve \cite{Cole-Schieve-2001},
and also took into account the Neumann boundary condition. 
The formulas obtained in the present paper enable us to get
exact numerical results for the energy density in a non-static cavity for an arbitrary
law of motion. Moreover, the structure of the formulas are meaningful. 
For instance, they exhibit that, apart from the law of motion, the evolution of the energy density
is completely determined by the total energy density in the static zone. In this
sense we can set up two cavities with different boundary conditions 
and different initial field states, but in such manner that both have the same energy density in
the static zone. In this situation we have (given the same law of motion for both
cavities) the same time evolution for the energy density. We pointed that an interesting
particular case occurs if we set up a non-mixed cavity with temperature
$0.34$ and a mixed cavity in the vacuum state. In this case we obtained
that the static and dynamical Casimir effect for the vacuum state in 
a situation where the Casimir force is repulsive (the mixed case)
are mimicked by another one for which the Casimir force is attractive
(the non-mixed case) but with a certain non vanishing temperature.

We obtained that the energy densities ${\cal T}_{\mbox{\footnotesize vac}}$, ${\cal T}_{\mbox{\footnotesize non-vac}}$ and
${\cal T}$ exhibit, in general, different structures. Wondering about the conditions under which they
would present the same behavior, we found that this particular situation 
can occur for diagonal states and laws of motion for which
the approximation given in Eq. (\ref{B-approx-A}) is valid. We showed that this 
condition is satisfied by the laws of motion investigated in Ref. \cite{estados-coerentes-2}, where
the same structure for these energy densities was obtained. On the other hand we showed that
if the condition (\ref{B-approx-A}) is not satisfied, these structures become different.

Focusing on the influence of the boundary conditions, we obtained that, for a same diagonal initial state 
and (arbitrary) law of motion, the energy density for the DD case is identical to that obtained for the NN case,
and the same occurs between ND and DN cases. This conclusion generalizes to any diagonal 
state the correspondent one obtained only for the vacuum case in Ref. \cite{Alves-Granhen-2008},
and also extends to the cavity problem the conclusion 
that the force acting on the moving boundary is invariant under the change $N\rightleftarrows D$
when the initial field state is invariant under time translations 
(in Refs. \cite{alves-granhen-lima-PRD-2008,Alves-Farina-Maia-Neto-JPA-2003}
this conclusion was obtained for the one-boundary case).
Investigating the thermal field, we obtained
specific formulas and concluded that, for a same temperature of the field state in the static zone, 
the energy density for the mixed case is bigger than the one obtained for the non-mixed case.
For non-diagonal states, which have the energy density
in the static zone non-invariant under time translation, 
we obtained that the energy density for the DD case in the dynamical situation is in general different from the one obtained for the NN case,
and the same is valid between DN and ND boundary conditions. 
However, we showed that in special situations the equalities can be recovered.
Investigating the coherent and Schrödinger cat states, we obtained
specific formulas for these cases and verified that the difference between DD and NN cases (and also
between ND and DN cases) is stored in ${\cal T}_{\left\langle \hat{a}\hat{a}\right\rangle_{\zeta}}$ 
part, whereas the ${\cal T}_{\left\langle \hat{a}^{\dag}\hat{a}\right\rangle_{\zeta}}$ part behaves like the energy density of
a diagonal state 	(all these conclusions are valid for any law of motion).
We applied our method of calculation
to the specific situation of a coherent state in a DD cavity with an 
oscillatory motion with small amplitude, 
verifying that the approximate result found in the literature 
for the energy density \cite{estados-coerentes-2} is in good agreement with the one obtained via the exact method presented here.
In addition, we also showed the correspondent results for NN, ND and DN cases.

We acknowledge F. D. Mazzitelli for useful discussions during the
Workshop ``60 Years of Casimir Effect''. We also acknowledge
V. V. Dodonov, A. L. C. Rego and M. A. Andreata for valuable discussions and suggestions. 
We are grateful to L. C. B. Crispino for careful reading 
of this paper. This work was supported by FAPESPA, CNPq and CAPES - Brazil.



\begin{thebibliography}{99}

\bibitem{Moore-1970} G. T. Moore, J Math. Phys. {\bf 11}, 2679 (1970).
%
\bibitem{Fulling-Davies-PRS-1976-I} S. A. Fulling and P. C. W. Davies, Proc. R. Soc. London, \textbf{A 348}, 393 (1976).
%
\bibitem{Fulling-Davies-PRS-1977-II} P. C. W. Davies and S. A. Fulling, Proc. R. Soc. London, \textbf{A 356}, 237 (1977).
%
\bibitem{trabalhos-pioneiros} B. S. DeWitt, Phys. Rep. \textbf{19}, 295 (1975);
P. C. W. Davies and S.A. Fulling, Proc. R. Soc. London, \textbf{A 354}, 59 (1977);
P. Candelas and D.J. Raine, J. Math. Phys. \textbf{17}, 2101 (1976);
P. Candelas and D. J. Raine, Proc. R. Soc. London, \textbf{A 354}, 79 (1977).
%
\bibitem{Vesnitskii-1972} A. I. Vesnitskii, Izv. Vyssh. Uchebn. Zaved. Radiofiz. \textbf{14}, 1432 (1971).
%
\bibitem{Law-PRL-94} C. K. Law, Phys. Rev. Lett. {\bf 73}, 1931 (1994).
%
\bibitem{solucoes-analiticas-exatas} Y. Wu, K. W. Chan, M. C. Chu, and P. T. Leung, Phys. Rev. A {\bf 59}, 1662
(1999); P. Wegrzyn, J. Phys. B {\bf 40}, 2621 (2007). 
%
\bibitem{Cole-Schieve-1995} C. K. Cole and W. C. Schieve, Phys. Rev. A {\bf 52}, 4405 (1995).
%
\bibitem{Dodonov-JMP-1993} V. V. Dodonov, A. B. Klimov, and D. E.
Nikonov, J. Math. Phys. {\bf 34}, 2742 (1993).
%
\bibitem{Dalvit-PRA-1998} D. A. R. Dalvit and F. D. Mazzitelli, Phys. Rev. A \textbf{57}, 2113 (1998).
%
\bibitem{perturbative-approach-one-boundary} L. H. Ford and A. Vilenkin, Phys. Rev. D \textbf{25},
2569 (1982); P. A. Maia Neto, J. Phys. A \textbf{27}, 2167 (1994);
P. A. Maia Neto and L. A. S. Machado, Phys. Rev. A \textbf{54}, 3420 (1996).
%
\bibitem{perturbative-approach-cavities} M. Razavy and J. Terning, Phys. Rev. D \textbf{31}, 307 (1985);
G. Calucci, J. Phys. A \textbf{25}, 3873 (1992); C. K. Law, Phys. Rev. A \textbf{49}, 433 (1994);
C. K. Law, Phys. Rev. A \textbf{51}, 2537 (1995); V. V. Dodonov and A. B. Klimov, Phys. Rev. A,
{\bf 53}, 2664 (1996); D. F. Mundarain and P. A. Maia Neto, Phys. Rev. A \textbf{57}, 1379 (1998).
%
\bibitem{temperatura-uma-fronteira} M. T. Jaekel and S. Reynaud, J. Phys. I (France) {\bf{3}}, 339 (1993);
M. T. Jaekel and S. Reynaud, Phys. Lett. A {\bf{172}}, 319 (1993);
L. A. S. Machado, P. A. Maia Neto, and C. Farina, Phys. Rev. D {\bf{66}}, 105016 (2002).
%
\bibitem{plunien-PRL-2000} G. Plunien, R. Schutzhold, and G. Soff, Phys. Rev. Lett. \textbf{84}, 1882 (2000).
%
\bibitem{alves-granhen-lima-PRD-2008} D. T. Alves, E. R. Granhen and M. G. Lima, Phys. Rev. D {\bf{77}}, 125001 (2008). 
%
\bibitem{temperatura-cavidade} J. Hui, S. Qing-Yun, and W. Jian-Sheng, Phys. Lett. A {\bf{268}}, 174 (2000);
R. Schutzhold, G. Plunien, and G. Soff, Phys. Rev. A {\bf{65}}, 043820 (2002);
G. Schaller, R. Schutzhold, G. Plunien, and G. Soff, Phys. Rev. A {\bf{66}}, 023812 (2002). 
%
\bibitem{Alves-Farina-Maia-Neto-JPA-2003}D. T. Alves, C. Farina, and P. A. Maia Neto, J. Phys. A \textbf{36}, 11333 (2003).
%
\bibitem{estados-coerentes} V. V. Dodonov, A. Klimov, and V. I. Man'ko, Phys. Lett. A
{\bf 149}, 225 (1990). 
%
\bibitem{estados-coerentes-2} M. A. Andreata and V. V. Dodonov, J. Phys. A \textbf{33}, 3209 (2000).
%
\bibitem{estados-coerentes-superpostos} V. V. Dodonov, M. A. Andreata, and S. S. Mizrahi, J. Opt. B: Quantum Semiclass.
Opt. {\bf 7} S468 (2005); D. A. R. Dalvit and P. A. Maia Neto, Phys. Rev. Lett {\bf 84}, 798 (2000).
%
\bibitem{cas-din-papel-conds-fronteira} M. Montazeri and M. F. Miri, 
Phys. Rev. A {\bf 71}, 063814 (2005); D. T. Alves, C. Farina, and E. R. Granhen, 
{\it ibid.} \textbf{73}, 063818 (2006); B. Mintz, C. Farina, P. A. M. Neto, and R. B. Rodrigues,
J. Phys. A \textbf{39}, 6559 (2006); J. Sarabadani and M. F. Miri, 
{\it ibid.} {\bf 75}, 055802 (2007). 
%
\bibitem{Alves-Granhen-2008} D. T. Alves and E. R. Granhen, Phys. Rev. A \textbf{77}, 
015808 (2008).
%
\bibitem{Cole-Schieve-2001} C. K. Cole and W. C. Schieve, Phys. Rev. A {\bf 64}, 023813 (2001).
%
\bibitem{Dalvit-JPA-2006} D. A. R. Dalvit, F. D. Mazzitelli, and O. Millán,
J. Phys. A \textbf{39}, 6261 (2006).
%
\bibitem{Boyer-AJP-2003} T. H. Boyer, Am. J. Phys. {\bf 71}, 990 (2003).
%
\bibitem{Boyer-PRA-1974} T. H. Boyer, Phys. Rev. A {\bf 9}, 2078 (1974).
%
\bibitem{Glauber} R. J. Glauber, Phys. Rev. {\bf{131}} 6, 2766 (1963);
R. J. Glauber, Phys. Rev. Letters. {\bf{10}} 84 (1963)  
%
\bibitem{Gato} E. Schrödinger, Naturwissensheften {\bf{23}}, 807 (1935);
C. C. Gerry and P. L. Knight, Am. J. Phys. {\bf{65}}, 964 (1997).
%
\bibitem{Jeong-2002} H. Jeong, A. P. Lund, and T. C. Ralph, Phys. Rev. A {\bf 72}, 013801 (2005).
%
\bibitem{Jeong-2005} H. Jeong and T. C. Ralph, quant-ph/0509137 (2005). 
%
\end{thebibliography}
\end{document}